\documentclass{aa}
\usepackage{graphicx}
\usepackage[T1]{fontenc}
\usepackage{ae,aecompl}
\usepackage{lscape}
\usepackage{txfonts}
\usepackage{natbib}
\bibpunct{(}{)}{;}{a}{}{,}
\begin{document}

\title{Active Region Moss}
\subtitle{Basic physical parameters and their temporal variation}

\author{D. Tripathi\inst{1}, H. E. Mason\inst{1}, G. Del Zanna\inst{1}, P. R. Young\inst{2, 3}}

\offprints{D.Tripathi@damtp.cam.ac.uk}

\institute{{DAMTP, University of Cambridge, Wilberforce Road, Cambridge CB3 0WA, UK} 
\and{Space Science Division, Naval Research Laboratory, Washington, DC 20375, USA} 
\and{George Mason University, 4400 University Drive, Fairfax, VA 22030, USA }}

\date{Received(16 December 2009); Accepted(10 May 2010)}

\abstract{Active region moss are transition region phenomena, first noted in the images recorded by the Transition Region and Coronal Explorer (TRACE) in $\lambda$171. Moss regions are thought to be the footpoints of hot loops (3-5~MK) seen in the core of active regions. These hot loops appear 'fuzzy' (unresolved). Therefore, it is difficult to study the physical plasma parameters in individual hot core loops and hence their heating mechanisms. Moss regions provide an excellent opportunity to study the physics of hot loops. In addition, they allow us to study the transition region dynamics in the footpoint regions.}{To derive the physical plasma parameters such as temperature, electron density, and filling factors in moss regions and to study their variation over a short (an hour) and a long time period (5 consecutive days).}{Primarily, we have analyzed spectroscopic observations recorded by the Extreme-ultraviolet Imaging Spectrometer (EIS) aboard Hinode. In addition we have used supplementary observations taken from TRACE and the X-Ray Telescope (XRT) aboard Hinode.}{The moss emission is strongest in the \ion{Fe}{xii} and \ion{Fe}{xiii} lines. Based on analyses using line ratios and emission measure we found that moss regions have a characteristic temperature of log~T[K]~=~ 6.2. The temperature structure in moss region remains almost identical from one region to another and it does not change with time. The electron densities measured at different locations in the moss regions using \ion{Fe}{xii} ratios are about 1-3~$\times$~10$^{10}$~cm$^{-3}$ and about 2-4~$\times$~10$^{9}$~cm$^{-3}$ using \ion{Fe}{xiii} and \ion{Fe}{xiv}. The densities in the moss regions are similar in different places and show very little variation over short and long time scales. The derived electron density substantially increased (by a factor of about 3-4 or even more in some cases) when a background subtraction was performed. The filling factor of the moss plasma can vary between 0.1-1 and the path length along which the emission originates is from a few 100 to a few 1000 kms long. By combining the observations recorded by TRACE, EIS and XRT, we find that the moss regions correspond to the footpoints of both hot and warm loops.}{}


\titlerunning{Active Region Moss}
\authorrunning{Durgesh Tripathi et al. }

\maketitle

\section {Introduction\label{intro}}

The high resolution images obtained by the Transition Region and
Coronal Explorer \citep[TRACE;][]{trace} revealed a new type of
emission called "moss". Moss regions are bright, finely textured,
mottled, low lying emission above the active region plage area. Moss 
regions are seen best in the TRACE images obtained at
\ion{Fe}{ix}/{x} $\lambda$171 \citep{schrijver, berger_moss}.  It has
been shown that moss regions are always observed in plage regions in the
vicinity of hot loops. These features are possibly the same phenomena
observed by \cite{prg} using the Normal Incidence X-ray Telescope,
where they found that many active regions were associated with
low-lying areas of intense emission resembling plage regions seen in
H$\alpha$ observations. Using observations from TRACE and the Soft
X-ray Telescope \citep[SXT;][]{sxt} it was suggested that the moss
regions correspond to the footpoint locations of hot loops which are
observed using X-ray images at 3-5~MK \citep{berger_moss, martens,
  antiochos}.

Active regions on the Sun primarily comprise two types of loops; the
loops seen in the hot and dense core of active regions in X-ray
observations at 2-3 MK (and higher) and the larger loops seen on the
periphery of active regions at ~1MK \citep[see
e.g.][]{delzanna&mason_2003}. The loops seen on the periphery of
active regions are termed "warm loops". With high spatial 
resolution instruments such as
TRACE, and the Extreme-ultraviolet Imaging Spectrometer
\cite[EIS;][]{eis} onboard Hinode \citep{hinode}, the warm loops 
seem to be spatially well resolved. Using TRACE and EIS observations 
the plasma parameters (such as electron density, temperature and
flows) 
in warm loops can be measured \citep[see e.g.,][] {warren_loop,
tripathi_2009}.

In contrast, the hot loops in the core of active regions appear
quite small, diffuse and difficult to resolve with present day
instrumentation. It has also been known for some time that the 
corona appears 'fuzzier' at higher temperatures. 
\citep{tripathi_2009} showed that this was not simply an
instrumental feature. This effect makes it very difficult to
resolve a single isolated loop structure in the core of an active
region. As a consequence, it is difficult to study the 
heating mechanism for individual hot loops in the core of 
active regions. A different approach is therefore required. 
Since it has been proposed that moss regions are the footpoints 
of hot loops, a detailed investigation of physical plasma 
parameters in moss regions and their variation with time should give 
some indication of the nature of the heating mechanism(s).

%%-------------------------
\begin{table}
\centering
\caption{Dates and raster start times of EIS data used in this study. \label{data}}
\begin{tabular}{cc}
\hline \hline
Date         							& Raster Start   \\
										&Times (UT)  \\
\hline
01-May-2007						& 11:53:13 \\
02-May-2007						& 05:06:11 \\
										& 18:31:20 \\
03-May-2007						& 14:01:52 \\
										& 14:21:12 \\
										& 14:40:31 \\
										& 14:59:51 \\
04-May-2007						& 06:37:17 \\
										& 06:56:36 \\
05-May-2007						& 05:24:09 \\
										& 07:25:56 \\
										& 07:45:16 \\
										& 08:04:35 \\	
\hline
\end{tabular}
\end{table}
%%-------------------------

In an earlier study \citep{tripathi_moss}, using a single dataset from
EIS, we measured the electron densities and magnetic field structures 
in moss regions. We found that the densities in moss regions were
higher than the surrounding regions in the active region and varied
within the range 10$^{10}$~-~10$^{10.5}$~cm$^{-3}$ from one moss
region to the other. In addition, we found that the moss regions were
primarily located in only one magnetic polarity region. In this paper, 
which is an extension of \cite{tripathi_moss}, we use observations 
recorded by EIS to study physical plasma parameters (such as electron
densities, temperatures, filling factors, and column depth) in different
moss regions within the same active region. In particular, we study 
the variation of these parameters over short
(one hour) and long (5 days) time periods. To the best of our
knowledge this is the first time that a spectroscopic study has been
carried out to study the variation of physical parameters in moss
regions over a short and a long period of time.

The rest of the paper is organized as follows. In section~\ref{obs} we
describe the observations used in this study. In section~\ref{techs}
we briefly discuss the spectroscopic techniques used in this paper. We also revisit the question of moss regions being the footpoints
of hot loops in section~\ref{moss_hot_loops} using data from TRACE,
EIS and the X-Ray Telescope \citep[XRT;][]{xrt}. We discuss the
thermal structure of moss regions in section~\ref{temp} followed by
a discussion of density, filling factors and column depth in 
section~\ref{dens}. We draw some conclusions in section~\ref{con}.

%%________________________________________
\begin{table}
\centering
\caption{Spectral lines (first column) from the study sequence
    \textsl'CAM\_ARTB\_CDS\_A' chosen to derive the physical
    parameters in moss regions. \label{lines}}
\begin{tabular}{lccc}
\hline \hline
Line~ID         	& Wavelength       			& log(N$_e$) Range   & 				log(T$_e$)   \\
&  ({\AA})         & (cm$^{-3}$)     & (K)      \\
\hline
\ion{Fe}{viii}		& 186.60		& $-$			&5.6	\\
\ion{Mg}{vii}     	& 278.39		& $-$		&5.8	\\
\ion{Mg}{vii}     	& 280.75		& 8.0$-$11.0		&5.8	\\
\ion{Si}{vii}		& 275.35		& $-$			&5.8	\\
\ion{Fe}{ix} 		& 188.50		& $-$			&6.0	\\	
\ion{Fe}{xi} 		& 188.23		& $-$			&6.1	\\
\ion{Si}{x}		& 258.37		& 8.0$-$9.7		&6.1	\\
\ion{Si}{x}		& 261.04		& $-$		&6.1	\\
\ion{Fe}{xii}		& 186.88		& 7.0$-$12.0       &6.1   	\\
\ion{Fe}{xii}    	& 195.12             	& $-$       &6.1  	\\
\ion{Fe}{xiii}		& 196.54		& 9.3$-$11.0		&6.2	\\
\ion{Fe}{xiii}    	& 202.02             	& $-$      		&6.2   	\\
\ion{Fe}{xiii}    	& 203.83             	& 8.0$-$10.5      	&6.2    	\\
\ion{Fe}{xiv}     	& 264.78             	& 8.0$-$11.0        &6.3   	\\
\ion{Fe}{xiv}     	& 274.20             	& $-$        &6.3   	\\
\ion{Fe}{xv}		& 284.16		& $-$ 			&6.4	\\
\hline

\end{tabular}
\tablefoot{The second column shows the central wavelengths, the third column shows the range of electron densities over which the ratios of lines are sensitive. Column four shows the peak formation temperature for the spectral lines.}
\end{table}

%%------
\section{Observations\label{obs}}
%%------

For this study, we have primarily used observations recorded by EIS
aboard Hinode. EIS has an off-axis paraboloid design with a focal
length of 1.9~meter and mirror diameter of 15~cm. It consists of a
multi-toroidal grating which disperses the spectrum onto two different
detectors covering 40~{\AA} each. The first detector covers the
wavelength range 170-210~{\AA} and the second covers 250-290~{\AA}
providing observations in a broad range of temperatures 
(log~T $\approx$~4.7-7.3). EIS has four slit/slot options available
(1\arcsec, 2\arcsec, 40\arcsec and 266\arcsec).  High spectral
resolution images can be obtained by rastering with a slit.

%%-------------
\begin{figure}   %% Figure 1
\centering
\includegraphics[width=0.35\textwidth]{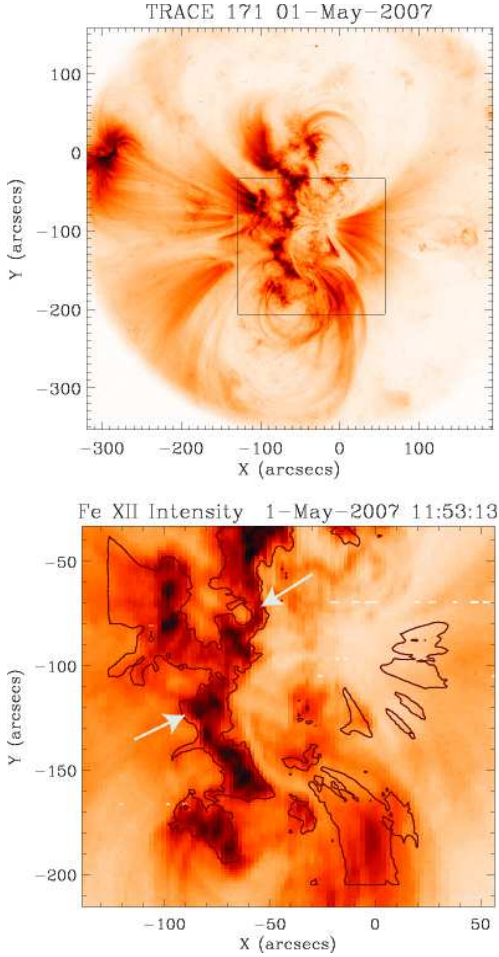}
\caption{Top panel: A TRACE image (plotted in a negative intensity
    scale) taken at 171~{\AA} showing the active region studied in
    this paper. The over-plotted rectangle shows the region which was
    rastered on May 01 using the 2$\arcsec$ slit of EIS. We note that
    a raster of this active region was obtained on 5 consecutive days with
    roughly the same coordinates of the boxed region.  Bottom panel:
    an EIS image in \ion{Fe}{xii}~$\lambda$195. The overplotted contours
    are from the TRACE intensity image. The vertical structure outlined
    with the contour in the middle of the image (also marked with
    arrows) is the moss region being discussed in the paper.
  \label{context}}
\end{figure}
%%------------

An active region \textsl{AR~10953}, which appeared on the east solar
limb on April 27, 2007, was observed by Hinode/EIS as it crossed the
visible solar disk. From May 1 till May 5th it was observed using the
study sequence \textsl{CAM\_ARTB\_CDS\_A} designed by the authors.
This study sequence takes about 20 minutes to raster a field of view
of the Sun of 200\arcsec~$\times$~200\arcsec~with an exposure time of
10~seconds using the 2\arcsec slit. It has 22 windows and is rich in
spectral lines, which allows us to derive the physical plasma parameters 
simultaneously at different temperatures. The top panel of 
Fig.~\ref{context} displays the active region imaged by TRACE in its 
171~{\AA} channel. The over-plotted box shows the portion of the 
active region which was scanned by EIS with its 2\arcsec~slit. The lower 
panel shows an EIS image in \ion{Fe}{xii}~$\lambda$195.12 line. The vertical 
structure in the middle of the image, outlined by the contour and also 
marked with arrows, locates the moss regions discussed throughout the paper.

The datasets comprise a couple of rasters each day, but these were 
not necessarily taken sequentially. On May 03, the study sequence was 
run four consecutive times with a cadence of 20 min each. This provides 
an excellent opportunity to study the physical characteristics of the 
moss over an hour. We have used these four datasets to 
study the variation of plasma parameters such as electron
temperature, density and filling factor. In addition we have taken one
raster each day from May 01 to May 05 to study the variation in moss
over a period of 5 days. In total we have analyzed 13 EIS datasets.
Table~\ref{data} contains dates and the start times of EIS rasters used
in this study.

Table~\ref{lines} provides the list of spectral lines (formed at log~T~=~5.6~-~6.5) used in this study. Four lines are affected by
blending, but for three of the lines the blending component can be
accurately estimated. \ion{Mg}{vii}~$\lambda$278.39 is blended with 
\ion{Si}{vii}~$\lambda$278.44 which has a fixed ratio relative to the 
unblended \ion{Si}{vii}~$\lambda$275.35 line and so can be easily 
evaluated \citep[see e.g.,][]{peter_eis}. \ion{Fe}{xiv}~$\lambda$274.20 is 
blended with \ion{Si}{vii}~$\lambda$274.18 which is generally much 
weaker. We estimate the \ion{Si}{vii}~$\lambda$274.18 contribution using 
\ion{Si}{vii}~$\lambda$275.35 which has its highest ratio of 0.25 in a density 
region of 10$^{10}$~cm$^{-3}$.

\ion{Fe}{xiii}~$\lambda$203.82 is partly blended with
\ion{Fe}{xii}~$\lambda$203.72 and the two components can be extracted by
simultaneously fitting two Gaussians to the observed spectral
feature. \ion{Fe}{viii}~$\lambda$186.60 is blended with \ion{Ca}{xiv}~$\lambda$186.61 
but it is not possible to estimate the blending
contribution using the available data. Since \ion{Ca}{xiv} is formed
at around $\log\,T=6.6$, it will only be significant in the core
of the active region, however this is where the moss regions are found
and so  \ion{Ca}{xiv} can be expected to be a significant contributor
to the \ion{Fe}{viii} line.

The \ion{Fe}{xii}~$\lambda$186 and $\lambda$195 lines are self blends.
For \ion{Fe}{xii}~$\lambda$186, we have fitted both of the lines with
one Gaussian and we have used both spectral lines in the CHIANTI~v6.0
\citep{chianti_v1, chianti_v6} model in the derivation of the density. The
\ion{Fe}{xii}~$\lambda$195.12 line is self-blended with the
\ion{Fe}{xii}~$\lambda$195.18 line \citep{fe12}. The ratio of these
two lines is sensitive to density. This blend can safely be ignored
for quiet solar active region conditions such as for quiescent active
region loops. However, the blend cannot be ignored while studying the
moss regions, where the electron density is well above
10$^{10}$~cm$^{-3}$ and the line at $\lambda$195.18 becomes $\sim$15\%
of the line at $\lambda$195.12. \cite{peter_dens} suggested that to deal with
the $\lambda $195.18 blend a two Gaussian fit can be performed,
where the stronger $\lambda$195.12 line has free parameters for the
centroid, width and intensity, while $\lambda $195.18 is forced to be
0.06~{\AA} towards the long wavelength side of $\lambda $195.12, and
to have the same line width as $\lambda $195.12. However, the
intensity of $\lambda$195.18 is free to vary. In this study, we have
used the technique suggested by \cite{peter_dens} to de-blend $\lambda
$195.12 from $\lambda$195.18.

%%-------------------------
\section{Spectroscopic techniques} \label{techs}
%%-------------------------
In order to derive physical parameters such as temperature, electron
density and filling factors, a number of different spectroscopic
techniques can be applied to EIS observations. For a review of
different spectroscopic techniques see e.g. \cite{dere_mason, mason}. 

The intensity of an optically thin emission line can be written as 

\begin{equation}\label{main}
I = 0.83~Ab(z) \int_{h} G(T_{\rm{e}}, N_{\rm{e}})~N_{e}^{2}~dh
\end{equation}

where Ab(z) is the elemental abundances, $T_{e}$ is the electron
temperature, and $N_{e}$ is the electron density. The factor 0.83 is
the ratio of protons to free electrons which is a constant for
temperatures above $10^5$~K.  G($T_{e}$, $N_{e}$)
is the \textit{contribution function} which contains all the relevant
atomic parameters for each line, in particular the ionization fraction
and excitation parameters and is defined as

\begin{equation}
G(T_{\rm{e}}, N_{\rm{e}}) = \frac{hc}{4{\pi} {\lambda_{i,j}}} \frac{A_{ji}}{N_{e}} \frac{N_{j}(X^{+m})}{N(X^{+m})}  \frac{N(X^{+m})}{N(X)}
\end{equation}

where i and j are the lower and upper levels, A$_{ji}$ is the
spontaneous transition probability, $\frac{N_{j}(X^{+m})} {N(X^{+m})}$
is the population of level j relative to the total $N(X^{+m})$ number
density of ion $X^{+m}$ and is a function of electron temperature and
density, $\frac{N(X^{+m})}{N(X)}$ is the ionization fraction which is
predominantly a function of temperature. The contribution functions 
for the emission lines considered here were computed with version~6 of the 
CHIANTI atomic database \citep{chianti_v6} using the CHIANTI ion balance 
calculations and the coronal abundances of \citep{coronal_abund}.

%%-----------------------
\subsection{Determination of electron temperature} \label{pott}
%%-----------------------

The solar plasma generally shows a continuous distribution of 
temperatures which is why such a broad range of ion species is seen in
the solar spectrum. The distribution is usually expressed as an emission
measure distribution that indicates the amount of plasma at each
temperature. In some cases solar plasma is found to be very close to
isothermal and an example is the quiet Sun plasma observed above the
limb \citep{feldman_1999}. A method that is very effective for establishing
if a plasma is isothermal is the so-called EM-loci method \citep[see
e.g. ][]{jordan,feldman_1999,em_loci}. In this method, the ratios of
observed intensities of different spectral lines with their corresponding 
contribution functions and abundances (i.e., I$_{obs}$/[Ab(z)~G(T$_{e}$, N$_{e}$)])
are plotted as a function of temperature. If the plasma is isothermal
along the line-of-sight (LOS) then all of the curves would cross at a single location
indicating a single temperature.

An indication of temperature can be obtained using emission lines from ions
with different ionization stages. As contribution functions are
generally sharply peaked functions in log temperature then ratios of two
contribution functions will be monotonic functions in temperature,
allowing observed intensity ratios to be converted to a temperature
estimate. The temperature is not physically meaningful if the plasma is
multithermal. However if the two ions are formed close to the dominant
emission temperature of the plasma then the ratio will accurately reveal those
locations with more high temperature plasma and those with more low
temperature plasma. For the present work we have used emission lines
from \ion{Fe}{xi} and \ion{Fe}{xiii} (Sect.~\ref{temp}).

As the moss studied here is found to be multithermal it is necessary
to perform an emission measure analysis to determine the temperature
distribution.  Here we follow the approach of \cite{pottasch} whereby
individual emission lines yield estimates of the emission measure
at the temperature of formation for each spectral line. By considering lines
formed over a wide range of temperatures, an emission measure
distribution can be determined. The method requires the contribution
function to be approximated by a simplified function such that $G$ is
defined to be a constant, $G_0$, over the temperature range $\log\,T_{\rm
  max}-0.15$ to $\log\,T_{\rm
  max}+0.15$ where $T_{\rm max}$ is the temperature where the
contribution function has its maximum. $G_0$ is evaluated as
\begin{equation}\label{em2}
G_0 = \frac{\int G(T_{e}, N_{e})~dT_{e}} {T_{\rm max} \times (10^{0.15} - 10^{-0.15})}.
\end{equation}
The expression for the line intensity, Eq.~\ref{main}, then becomes
\begin{equation}
I=0.83Ab(z) G_0 \int N_{\rm e}^2 dh.
\end{equation}
The emission measure for the emission line is then defined as
\begin{equation}\label{em-def}
EM= \int N_{\rm e}^2 dh
\end{equation}
and so
\begin{equation}\label{em3}
EM = {I_{\rm obs} \over 0.83 Ab(z) G_0 }
\end{equation}
thus the emission measure is defined entirely by the observed line
intensity, the element abundances and the atomic parameters contained
in $G_0$.

An IDL routine called \emph{integral\_calc.pro} available in the
CHIANTI software distribution is used here to compute the quantity $G_0$.

%%-----------------------
\subsection{Determination of electron density, filling factor and column depth}
%%-----------------------
The electron density of an astrophysical plasma can be derived by
measuring two emission lines of the same ion that have different
sensitivities to the plasma density, the ratio yielding a direct
estimate of the density \citep[e.g.][]{mason}. This method is
independent of the emitting volume, element abundances or ionization
state of the plasma, and depends solely on the atomic population
processes within the ion.

%%-------------------------
\begin{figure} %% Figure 2
\centering
\includegraphics[width=0.4\textwidth]{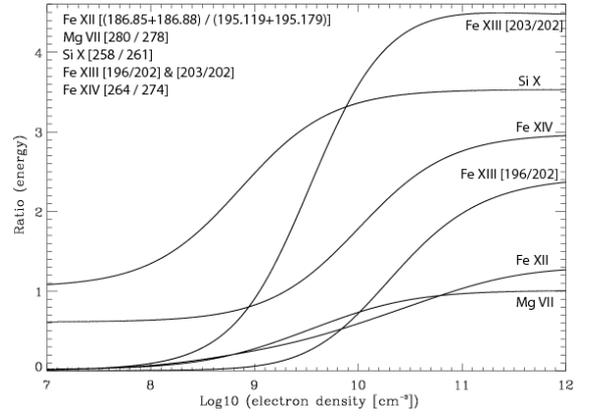}
\caption{CHIANTI, v6.0, theoretical intensity ratios with respect to electron density 
for the spectral line ratios used in this paper. The spectral lines are labelled on the plot.\label{chianti}}
\end{figure}
%%-------------------------

EIS provides access to a number of line ratio density diagnostics formed at
different temperatures and
Table~\ref{lines} lists the diagnostics observed with the observation study
\textsl{CAM\_ARTB\_CDS\_A}. The theoretical variations of the line
ratios with density are derived using version~6 of the CHIANTI
database (Dere et al. 2009) and the curves are shown in
Fig.~\ref{chianti}. 

The density can be used to derive the filling factor of the
plasma. If we assume that the density is constant within the emitting
volume for the ion then the emission measure (Eq.~\ref{em-def}) can be
written as $N_{\rm e}^2 h$ where $h$ is the column depth of the
emitting plasma. Rearranging Eq.~\ref{em3} then gives
\begin{equation}\label{coldepth}
h = {I_{\rm obs} \over 0.83 Ab(z) G_0 N_{\rm e}^2 }.
\end{equation}
By inspecting images of the emitting plasma, it is possible to
determine the apparent column depth of the plasma, $h_{\rm
  app}$. In the present case this is done by studying images of the
moss as the active region approaches the limb (Sect.~\ref{dens}).
That is when the radial
dimension of the moss is almost perpendicular to the line of sight 
and so its depth can be measured visually. The ratio of the
spectroscopically derived column depth, $h$, to $h_{\rm app}$ then
yields a value for the filling factor, $\phi$, of the plasma. i.e.,
\begin{equation}\label{fill}
\phi = { EM \over  N_{\rm e}^2 h_{\rm app} }.
\end{equation}
$\phi$ essentially measures the fraction of the observed plasma
volume that is actually emitting the emission line under study. Values
less than one imply that the volume is not completely filled with
emitting plasma. 

%%----------------------------------------------------
\section{Active region moss and hot loops} \label{moss_hot_loops}
%%----------------------------------------------------

%%-----------------------
\begin{figure}   %% Figure 3
\centering
\includegraphics[width=0.45\textwidth]{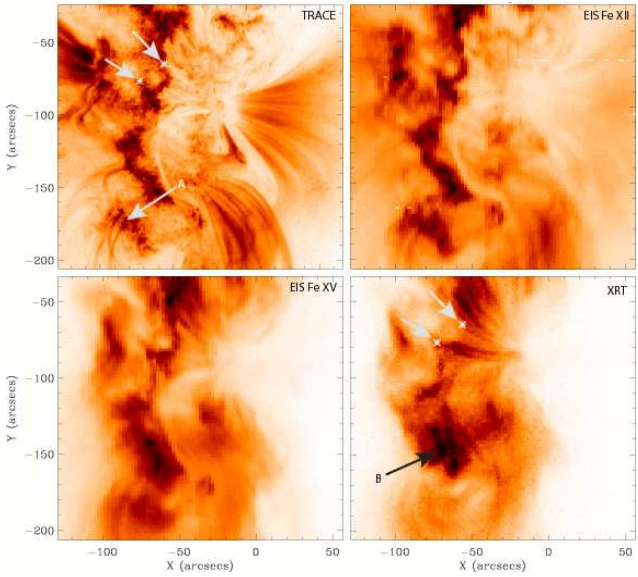}
\caption{Co-aligned TRACE~$\lambda$171 (top left), EIS
  \ion{Fe}{xii}~$\lambda$195 (top right), EIS
  \ion{Fe}{xv}~$\lambda$284 (bottom left) and XRT~Al\_poly filter
  images (bottom right) taken on May 01, 2007.  The images are shown
  in negative intensity. The two stars indicated by two arrows in the
  TRACE image show moss regions. The two stars in the XRT image are
  located at the same position as those in the TRACE image showing the
  footpoints of hot loops. The arrow 'A' locates a couple of 1MK
  loops rooted in moss regions. Arrow 'B' in the XRT image shows the
  high temperature fuzzy emission in the moss
  regions.}\label{cross_cor_image}
\end{figure}
%%-----------------------

Based on the observations recorded from TRACE and SXT and using
analytical calculations it has been proposed that the moss regions are
the footpoints of the hot loops seen in the SXT images taken at
3-5~MK. However, we note that the spatial resolution of TRACE is a
factor of 2.5 better than the high resolution SXT images. The X-ray 
images recorded by XRT aboard Hinode are of comparable
resolution to that of TRACE images (1 arcsec per pixel). In addition, spectral 
images obtained using EIS provide further information at intermediate
temperatures. 
Hence, we have revisited this relationship question in this paper using TRACE, 
XRT and EIS observations.

In order to compare the observations taken from XRT, TRACE and 
EIS, a coalignment of the images was performed. It is known that images 
taken using the two CCDs of EIS are shifted with respect to each other 
\citep{peter_artb}. To coalign the EIS spectral images obtained from the 
two detectors, we cross-correlated images obtained in \ion{Fe}{xii}~$\lambda$195 
and \ion{Si}{x}~$\lambda$261. Since the 
peak formation temperature of these two lines are the same, they reveal 
the same structures. The TRACE~$\lambda$171 and XRT~Al\_poly images were then cross-correlated with 
the images obtained in \ion{Fe}{xii}~$\lambda$195 and  \ion{Fe}{xv}~$\lambda$284 respectively.

Figure~\ref{cross_cor_image} displays co-aligned images recorded from
TRACE~$\lambda$171 (top left panel), EIS \ion{Fe}{xii} (top right),
EIS \ion{Fe}{xv} (bottom left) and XRT (using the Al\_poly
filter) (bottom right panel). The images are displayed in a negative
intensity scale. The bright
moss regions can be seen as dark regions located in the left half of
the top left image, as also shown in the bottom panel of Fig.~\ref{context}. 
We have plotted two asterisks on the TRACE image,
shown by two arrows in the top left image, locating moss regions. The
two asterisks in the bottom right image (also shown by two arrows)
correspond to the same locations as those in the top right image. This
clearly demonstrates that those moss regions are essentially located
at the footpoints of the hot loops as deduced previously
  \citep[see e.g.,][]{antiochos}. The arrow labelled as 'A' in the
top left panel locates an 1MK loop located in
the moss regions and coexistent with high temperature loops seen in EIS~\ion{Fe}{xv} and XRT images. Therefore, it appears reasonable to deduce that the
moss regions are not just the footpoints of hot loops, rather there
are warm loops at 1MK which are also rooted in the moss regions.
However, we cannot rule out the possibility that these warm
  loops are those which are cooling down to 1MK from a temperature of
  2-3MK i.e., from XRT temperatures to TRACE.   
  
  The arrow labelled as 'B' in the XRT
image shows hot fuzzy emission, which is located over moss regions
when compared to the top left image. The loop structures are not clear
and it is difficult to deduce if these moss regions are the footpoints
of loops.

%%-----------------------
\section{Thermal structure of moss}\label{temp}
%%-----------------------
%%---------------------
\begin{figure}  %% Figure 4
\centering
\includegraphics[width=0.5\textwidth]{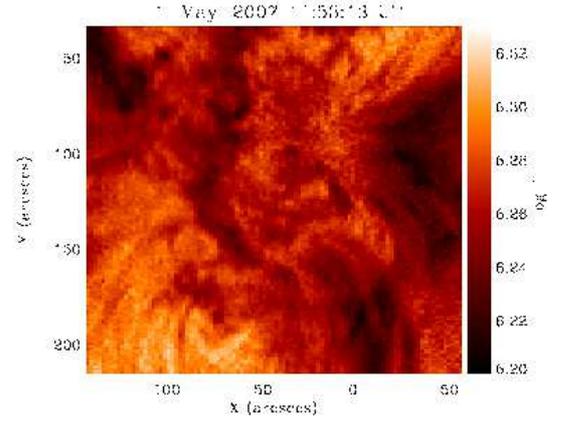}
\caption{A temperature map derived using intensity ratios of
  \ion{Fe}{xi}~$\lambda$188 and \ion{Fe}{xiii}~$\lambda$202. We used
  the ionization fraction from CHIANTI~v6.0. \label{temp_map}}
\end{figure}
%%---------------------

%%----------------------
\begin{figure}   %% Figure 5
\centering
%\sidecaption
\includegraphics[width=0.4\textwidth]{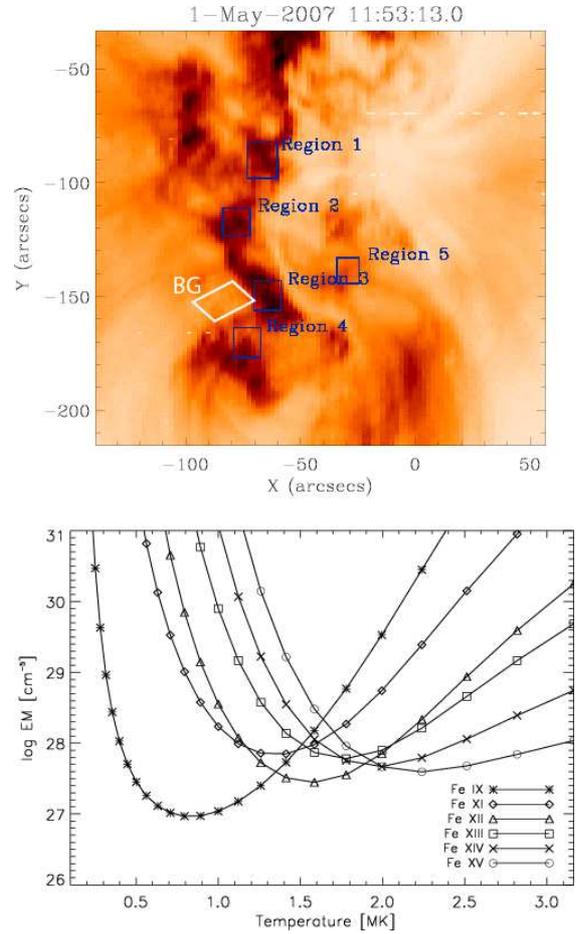}
\caption{Top panel: an \ion{Fe}{xii} image from an EIS raster. The
  overplotted boxes show the regions which were used for
  deriving plasma parameters. BG is the region which was used to
  subtract the background emission in section~\ref{fill_factor}. Bottom
  panel: an EM-loci plot for region 1 using the ionization balance from
  CHIANTI~v6.0 and the coronal abundances of \cite{coronal_abund}. The
  meaning of the different symbols are shown in the figure.\label{em_loci}}
\end{figure}
%%----------------------

%%-------------------------
\begin{figure}  %% Figure 6
\centering
\includegraphics[width=0.45\textwidth]{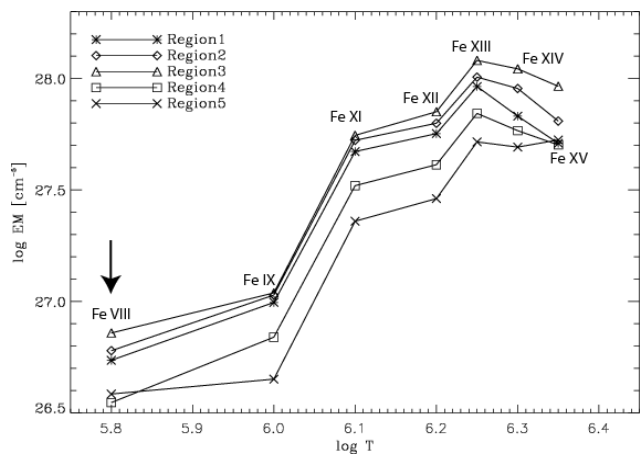}
\caption{An average emission measure plot for the five regions shown in 
the top panel of Fig~\ref{em_loci}. CHIANTI~v6.0 ionization 
equilibrium and the coronal abundances of \cite{coronal_abund} wewre
used. The EM derived for \ion{Fe}{viii} is an upper limit and is marked with an arrow.\label{emission_measure}}
\end{figure}
%%-------------------------

The moss regions were originally noted in the images recorded using the 
171~{\AA} channel of TRACE, which primarily observes solar
transition region plasma at a temperature of $\sim$~1~MK.
\cite{fletcher_moss}, using a DEM study of an observation taken from
SoHO/CDS, showed that the plasma in moss regions was multi-thermal. 
Recent studies using EIS data \citep[see e.g.,][]{warren_moss, tripathi_moss} 
confirmed that moss regions are seen not only at 1~MK but in a range of 
temperatures. Therefore, in order to understand the physics of moss regions, it is important to
understand the thermal structure of moss and its temporal variation.

Fig.~\ref{temp_map} displays a temperature map of the active region
rastered on May 01, 2007 which was derived using intensity ratios of
\ion{Fe}{xi}~$\lambda$188.2 and \ion{Fe}{xiii}~$\lambda$202.0 using the
ionization fraction from CHIANTI~v.6.0 and the coronal abundances of
\cite{coronal_abund}. The temperature map shows that most of the moss regions 
(corresponding to the contoured regions in the bottom panel of Fig.~\ref{context}) is
within the temperature range of log~T~=~6.2~-~6.3. This basically
reflects the fact that by taking ratios, we are measuring a
temperature common to the \textit{contribution functions} of the two
spectral lines. From the figure, however, it is evident that the moss
regions are at a temperature of log~T~=~6.2. In addition, we find that
moss regions are cooler than some of the surrounding regions. Indeed,
for those regions, we found the existence of hot emission by
investigating the spectral images obtained in \ion{Fe}{xiv} and
\ion{Fe}{xv} lines.

The bottom panel in Fig.~\ref{em_loci} shows the 'EM-loci' plots for one
region (region 1) of the moss, labelled in the top panel. In order to
compute the EM-loci plots, we have only used the spectral lines of
iron, so that we can rule out any effects of abundance variations on the
relative magnitudes of the emission measures obtained for different
spectral lines. As can be deduced from the figure, the plasma along
the LOS is multi-thermal. Most of the emission, however, is within the
temperature range 1.2~MK to 1.8~MK. The peak of emission measure is at
around log~T~=~6.2, suggesting a similar temperature for the moss to that derived from
the line ratios. For all the five regions marked in Fig.~\ref{em_loci}
(top panel), the crossing point of the curves are very similar.
However, the magnitude of the emission measure crossing point 
is different for different regions.
In order to check the variation in the thermal structure of the moss
regions, we generated EM-loci plots of all 13 datasets listed in Table~\ref{data}. 
The crossing points of the EM curves were similar to that shown in the bottom 
panel of Fig.~\ref{em_loci}, however the magnitude of the emission measure 
did vary from region to region. We also considered each raster for five 
consecutive days and traced a specific region in all of the rasters. 
The EM-loci plots obtained for each region for all five days showed remarkable
similarities in terms of the crossing points of the curves. 
Therefore, we conclude 
that the thermal structure of the moss region remains fairly constant, 
at least for the active region studied in this paper and that most of the plasma 
in the moss region is in the temperature range 1.2~MK to 1.8~MK. Hence, 
the EM-loci plot presented in Fig.~\ref{em_loci} can be taken as typical for 
all regions of moss in this study.

The EM-loci analysis indicates that the plasma along the LOS in the moss
regions is multi-thermal. Therefore, in order to get a proper thermal
structure, we need to perform an EM analysis. For this purpose we have
used the Pottasch method as described in subsection~\ref{pott}.
Figure~\ref{emission_measure} shows a plot of the average EM
for all of the five regions shown in the top panel of
Fig.~\ref{em_loci}. The EM was calculated using
ionization fraction from CHIANTI~v6.0 and the coronal abundances of
\cite{coronal_abund}. In addition, for \ion{Fe}{viii} and \ion{Fe}{ix}
we have used densities derived using \ion{Mg}{vii} (formed at a
similar temperature), for \ion{Fe}{xi}
and \ion{Fe}{xii} we have used densities derived from \ion{Fe}{xii}
and for \ion{Fe}{xiii} and \ion{Fe}{xiv} we have used 
densities dervided from diagnostic line ratios within those ions. 
As is evident from the plot, 
most of the emission in moss regions is observed in \ion{Fe}{xiii}
in all cases.
The emission starts to decrease in \ion{Fe}{xiv} and \ion{Fe}{xv}.
From the plot it appears that \ion{Fe}{xiii} is the turning point of
the emission measure curve. It is likely that the emission seen in
\ion{Fe}{xiii} is not just from the moss emission, but is possibly
contaminated with emission from hot loops which are seen in 
\ion{Fe}{xiv} and \ion{Fe}{xv}. The plot shows very little difference in the
emission seen in \ion{Fe}{viii} and \ion{Fe}{ix}. This could be due to
the fact that the \ion{Fe}{viii}~$\lambda$186.6 line used in this
study is blended with another line, \ion{Ca}{xiv}~$\lambda$186.61 
formed at log~T~=~6.4, and could therefore be contaminated with 
some emission from hot loops overlying the moss regions. The plot shown in
Fig.~\ref{emission_measure} suggests that a temperature somewhat close
to the formation temperature of \ion{Fe}{xi} and \ion{Fe}{xiii}
(log~T~=~6.1-6.3) is the characteristic temperature of the moss for
this active region.

%%------------------------------------------------------------------------
\begin{figure*}  % Figure 7
\centering
\sidecaption
\includegraphics[width=0.8\textwidth]{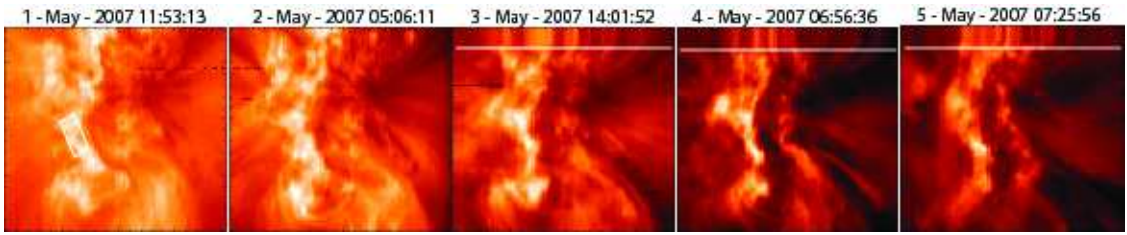}
\caption{Coaligned intensity images obtained in
  \ion{Fe}{xii}~$\lambda$195 for five consecutive days. The box shown
  in the left most image is chosen to perform the emission meassure
  analysis.\label{int_five_days}}
\end{figure*}
%%--------------

To study the variation of thermal structure in the moss
  regions over a period of five days, we have considered one raster
  every day and performed an emission measure analysis in a specific
  region. For the coalignment we cross-correlated the rasters obtained on 
  consecutive days. Fig.~\ref{int_five_days} shows the co-aligned 
  intensity images for five
  consecutive days obtained in \ion{Fe}{xii}~$\lambda$195. The data
  above the white lines in the last three images show the artifacts
  introduced due to cross-correlation and interpolation. We believe
  that we have achieved the co-alignment within a few arcsec. The
  overall structure of the active region stays almost the same. 
Fig.~\ref{five_days_em} shows
  emission measure as a function of temperature for the boxed region
  shown in the left image in Fig.~\ref{int_five_days}. As discussed earlier 
  and shown in Fig.~\ref{emission_measure} most of the emission in the moss
  region comes from \ion{Fe}{xiii} and the emission starts to decrease in
  \ion{Fe}{xiv}. It can be easily seen from the plot that the
  average emission measure for the boxed region remains fairly 
  constant over the five day period for all five spectral lines.
  This suggests that the thermal structure in the moss region does not
  change significantly with its temporal evolution.

%%-----------
\section{Densities and filling factors in moss} \label{dens}
%%-----------
\subsection{Densities in moss regions}
\begin{figure}  %% Figure 9
\centering
\includegraphics[width=0.45\textwidth]{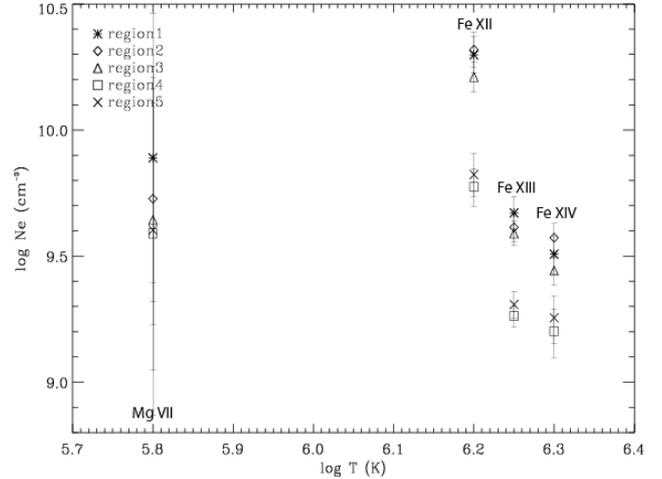}
\caption{Electron density measured using \ion{Mg}{vii}, \ion{Fe}{xii},
  \ion{Fe}{xiii} and \ion{Fe}{xiv} for the five different regions
  marked in the top panel in Fig~\ref{em_loci}. \label{density_may1}}
\end{figure}

Figure~\ref{density_may1} gives the densities measured in five
different moss regions (shown in the top panel of Fig.~\ref{em_loci})
simultaneously at different temperatures using the spectral lines
\ion{Mg}{vii} (log~T~=~5.8), \ion{Fe}{xii} (log~T~=~6.2),
\ion{Fe}{xiii} (log~T~=~6.25), \ion{Fe}{xiv} (log~T~=~6.3). The
uncertainties on the densities are calculated using 1-sigma errors in the
intensities derived from a Gaussian fitting of the spectral lines and
the photon statistics. In addition the errors for the derived electron
densities from the theoretical CHIANTI curves are estimated. These are
larger when the curves approach their high and low density
limits. The plot demonstrates that the electron density in each
  moss region falls off with temperature except for that derived from
  \ion{Mg}{vii}. However, we note the large error bars on the
  \ion{Mg}{vii} densities. These large errors are due to the fact
  that the two \ion{Mg}{vii} lines in EIS active region spectra are
  very weak \citep[see e.g.,][]{peter_artb}. It is also worth pointing out 
that the densities obtained using \ion{Fe}{xii} lines are much
  higher that those obtained using \ion{Fe}{xiii} and \ion{Fe}{xiv}.
  The decrease in the densities with temperature seen in the figure
  is anticipated if we assume a constant pressure in a
  given moss region. However, considering the peak formation temperature
  for each line and the corresponding derived densities we find that
  the pressure for \ion{Fe}{xii} is substantially higher than those
  for \ion{Fe}{xiii} and \ion{Fe}{xiv}.

%%---------------
\begin{figure}     %% Figure 8
\centering
\includegraphics[width=0.45\textwidth]{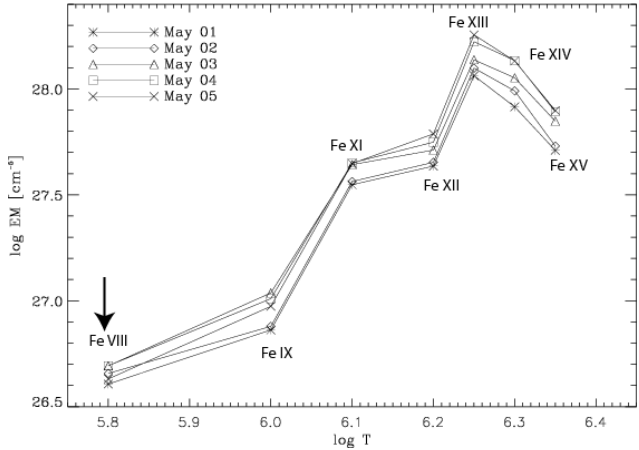}
\caption{An emission measure curve (obtained using the Pottasch method) for
  five days for the region shown in Fig.~\ref{int_five_days}.
  CHIANTI~v6.0 ionization equilibrium and the coronal abundances of
  \cite{coronal_abund} were used. The EM for \ion{Fe}{viii} is an upper limit and is marked with an arrow. \label{five_days_em}}
\end{figure}
%%---------------

%%-------------------------
%%-------------------------

%%----------------
\begin{figure}   %% Figure 10
%\sidecaption
\centering
\includegraphics[width=0.45\textwidth]{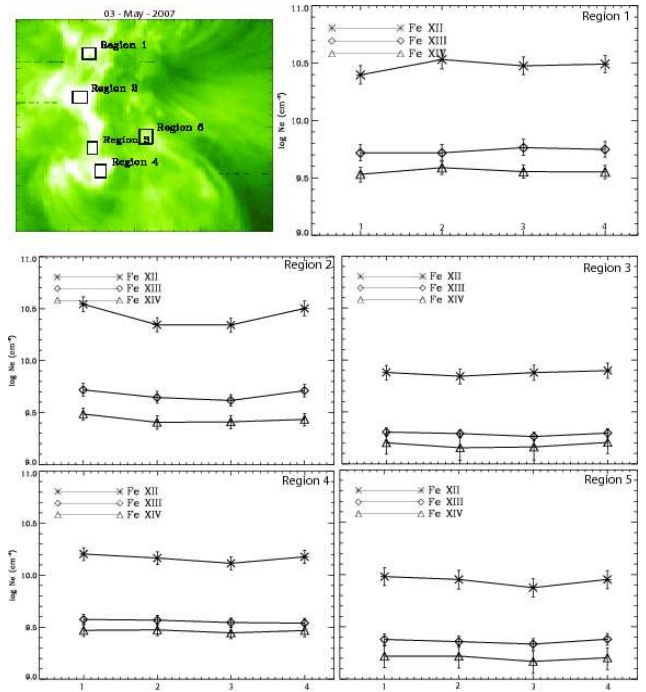}
\caption{The electron density variation in the moss regions over an hour
  derived from four consecutive EIS rasters. \label{dens_anhour}}
\end{figure}
%%--------------
%%-------------------------
\begin{figure}   %% Figure 11
\centering
\includegraphics[width=0.45\textwidth]{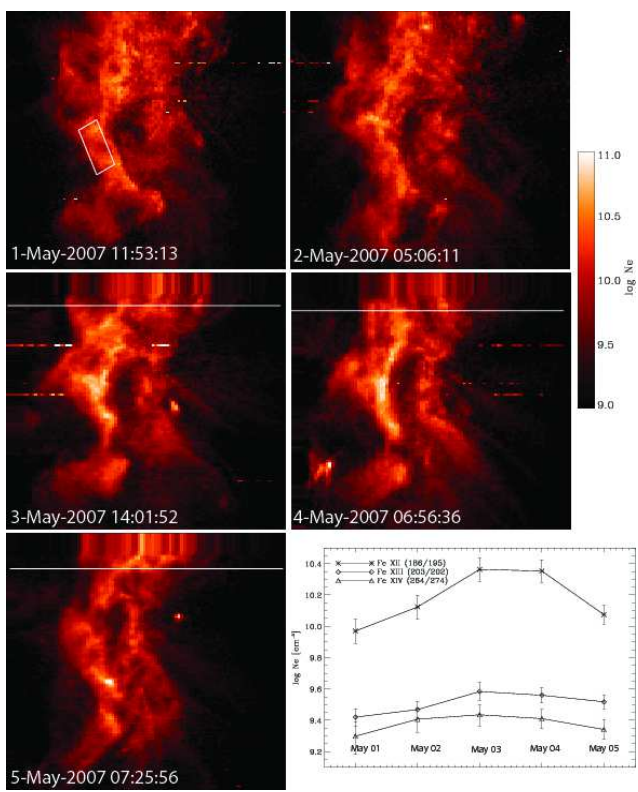}
\caption{Electron density maps obtained using coaligned intensity maps shown in
  Fig.~\ref{int_five_days}. The bottom right panel displays the
  electron density
  variation over a period of five days measured using \ion{Fe}{xii},
  \ion{Fe}{xiii} and \ion{Fe}{xiv} for the boxed region shown in the
  top left panel. \label{dens_five_days}}
\end{figure}
%%---------------

We have studied the short term and long term temporal variations of electron densities in the moss regions. For this purpose we have considered four consecutive rasters taken 20 minutes apart on May 03, 2007. Fig.~\ref{dens_anhour} displays the variation of electron densities derived using \ion{Fe}{xii}, \ion{Fe}{xiii}, and \ion{Fe}{xiv} for the five regions shown in the top left panel of the figure. Electron density values for different rasters are plotted with different symbols in each plot. The error bars are estimated as in Fig.~\ref{density_may1}. The uncertainties in the densities obtained using \ion{Mg}{vii} are very large, so we have omitted \ion{Mg}{vii} from the plot. The density falls off with temperature in a similar way to the plot shown in Fig.~\ref{density_may1}. It is also clear from the plot that the densities obtained using \ion{Fe}{xii} are consistently higher than those
  derived using \ion{Fe}{xiii} and \ion{Fe}{xiv}. The densities
  derived using \ion{Fe}{xiii} are also always larger than those by
  \ion{Fe}{xiv}. However, the small difference between the densities obtained
  from \ion{Fe}{xiii} and \ion{Fe}{xiv} could be real indicating
constant pressure. The plots clearly demonstrate 
  that there is almost no change in electron density over an hour at all three temperatures.

To study the evolution of electron density in the moss regions over a
  period of five days, we have considered the boxed region
  corresponding to the one shown in the left panel of
  Fig.~\ref{int_five_days}. The density maps corresponding to the
  intensity maps shown in Fig.~\ref{int_five_days} are shown in
  Fig.~\ref{dens_five_days}. The box in the top left image in
  Fig.~\ref{dens_five_days} corresponds to the region boxed in the
  intensity image shown in Fig.~\ref{int_five_days}. As with the
  intensity images, the data above the white lines in the last three
  density maps shows an artifact introduced due to cross-correlation.
  As can be seen from the figure, the overall density structure stays
  fairly similar as does the intensity structure (see
  Fig.~\ref{int_five_days}) with just a small enhancement in the
  center of the moss region. To show this quantitatively, we have studied
  the density variation in the boxed region shown in the top left image. 
The bottom right panel in Fig.~\ref{dens_five_days} displays
  the density variation showing that the electron density increases
  from May 1 to May 3 and then decreases on May 4, which is quite
  pronounced in \ion{Fe}{xii} and very slightly in \ion{Fe}{xiii} and \ion{Fe}{xiv}. 
  Except for this enhancement, we find that
  the electron density remains fairly similar in the moss regions and
  does not show much variation in time.  Although the reason for this
  enhancement in densities is not clear to us, we anticipate that this
  could be due to small scale dynamic activity taking place in the core
  of active regions seen in an XRT movie for this region. 
Using Coronal Diagnostic Spectrometer
  \citep[CDS;][]{cds} and Michelson Doppler Imager
  \citep[MDI;][]{mdi} data, \cite{soho17, helen_book} showed that 
localized enhancements in electron densities were correlated
  with emerging and canceling flux regions. We also note that canceling 
  flux regions are frequently observed near the polarity inversion line 
  \citep[see e.g., ][]{thesis}. However, further investigation of this is needed.

In all of the measurements so far presented in this paper and
  those results presented in other papers for high density regions
 (that is densities greater than
  10$^9$~cm$^{-3}$) cf
   \cite{tripathi_moss, warren_moss, peter_dens,
    brendan}, the densities measured using \ion{Fe}{xii} are
  reported to be larger than those obtained from \ion{Fe}{xiii} and
  \ion{Fe}{xiv}. It is worth mentioning here that if the
  \ion{Fe}{xii}~$\lambda$186.8 line were blended and we lower the
  intensity by 20\%, the electron densities obtained using
  \ion{Fe}{xii} would become consistent with those obtained using
  \ion{Fe}{xiii} and \ion{Fe}{xiv}. However, we cannot at present
  explain these discrepancies and so we leave this as an open question.

%%-----------------------------------------------
\subsection{Filling factors in moss regions}\label{fill_factor}
\begin{figure}    %% Figure 12
\centering
\includegraphics[width=0.4\textwidth]{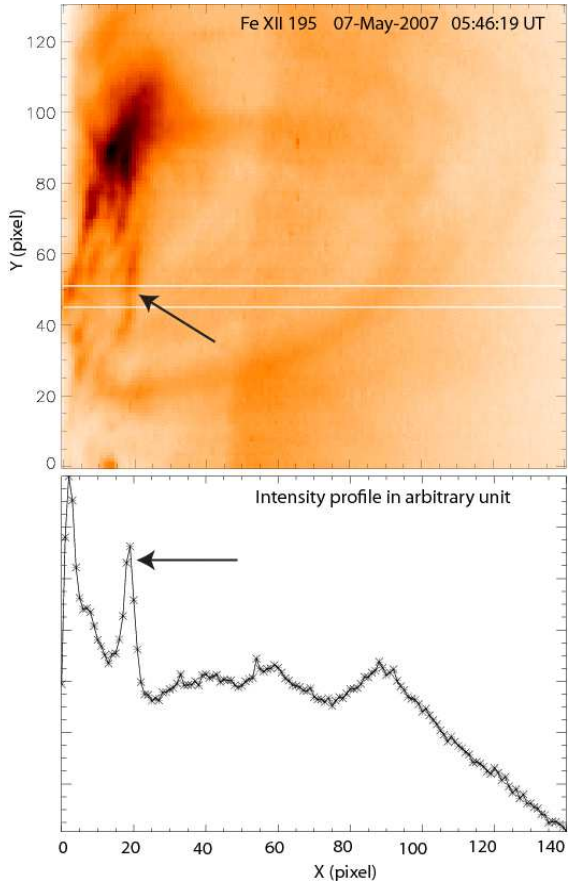}
\caption{Top panel: Negative intensity image for
  \ion{Fe}{XII}~$\lambda$195 on 07-May-2007. Bottom Panel:
  the intensity profile for \ion{Fe}{XII}~$\lambda$195
  between the two lines marked in the top panel. The
  arrow in the top panel marks the moss region and in the bottom panel
  marks the intensity enhancement due to the moss region.\label{thickness}}
\end{figure}

%%-----------------------------------------------
Equation~\ref{fill} gives the expression for deriving the filling
  factor from the emission measure, density and apparent column depth. The
  emission measure and density are derived directly from the
  spectroscopic data as described in the previous sections. To
  estimate the apparent column depth we follow the method of \cite{martens} and
  study images of the moss at the solar limb. The active region was
  observed close to the limb with EIS on 2007 May 7 and radial intensity
  profiles cutting through a particular moss region were studied in lines of
  \ion{Fe}{xii}, \ion{Fe}{xiii} and \ion{Fe}{xiv}. A sample
  intensity profile from \ion{Fe}{xii} $\lambda$195.12 is shown in
  Fig.~\ref{thickness} where a distinctive spike in emission
  corresponding to the moss region can be seen. We interpret the width
  of this spike to be the column depth of the moss, which is found to
  be 6\arcsec\ ($\sim$4000~km) in this case, in good agreement with the 
  results of \cite{martens}.

%%------------------------------------------------------------------------
%%-----------------------------------------------------------------------

%%---------------------------------------------------------------------
\begin{table*}
  \caption{Density (N$_e$), column depth (h) and filling factor ($\phi$) measurements for five 
  moss regions shown in Fig.~\ref{em_loci}. \label{table:fill}}
  \centering
\begin{tabular}{c|c|c|c|c|c|c|c|c|c}
\hline\hline
			&\multicolumn{9}{c}{\textbf{Before background subtraction}}\\
\cline{2-10}
 			&   \multicolumn{3}{c|}{\ion{Fe}{xii}} & \multicolumn{3}{c|}{\ion{Fe}{xiii}} & \multicolumn{3}{c}{\ion{Fe}{xiv}}\\
\cline{2-4}
\cline{5-7}
\cline{8-10}
          		& log N$_e$      				& h 					&  $\phi$     	&  log N$_e$ 				& h 							&  $\phi$    	&   log N$_e$      			&  h							&$\phi$ \\
          		& (cm$^{-3}$)				&  (km)      			& 	   		& (cm$^{-3}$)				& (km)       					&   			&   (cm$^{-3}$) 			&   (km)      					&  \\
\hline

Region1		& 10.32$^{10.39}_{10.25}$	& 140$^{100}_{200}$	& 0.04 		& 9.73$^{9.76}_{9.70}$		& 3000$^{2700}_{3300}$		& 0.8		& 9.54$^{9.60}_{9.48}$	& 5600$^{4400}_{7100}$	 	& 1.4\\
Region2		& 10.32$^{10.39}_{10.26}$	& 160$^{120}_{210}$	& 0.04		& 9.68$^{9.71}_{9.66}$		& 4200$^{3800}_{4300}$		& 1.0		& 9.57$^{9.62}_{9.51}$	& 6600$^{5400}_{8400}$	 	& 1.6\\
Region3		& 10.22$^{10.28}_{10.15}$	& 280$^{220}_{390}$	& 0.07		& 9.63$^{9.65}_{9.60}$		& 6400$^{6000}_{6900}$		& 1.6		& 9.45$^{9.50}_{9.40}$	& 14000$^{11000}_{17000}$ 	& 3.5\\
Region4 	&  9.79$^{9.86}_{9.72}$		& 1200$^{800}_{1500}$& 0.3		& 9.33$^{9.36}_{9.30}$		& 14000$^{13000}_{15000}$	& 3.5		& 9.21$^{9.29}_{9.12}$	& 22000$^{16000}_{33000}$ 	& 5.6\\
Region5		&  9.83$^{9.91}_{9.74}$		&  680$^{500}_{1000}$& 0.2		& 9.37$^{9.41}_{9.34}$		&  8800$^{7700}_{9500}$		& 2.2		& 9.25$^{9.33}_{9.15}$	& 16000$^{11000}_{24000}$ 	& 3.9\\
\hline
&\multicolumn{9}{c}{\textbf{After background subtraction}}\\
\cline{2-4}
\cline{5-7}
\cline{8-10}
Region1		& 11.00$^{11.08}_{10.93}$	& 8$^{6}_{10}$		& 0.002		& 10.58$^{10.63}_{10.54}$	& 130$^{100}_{150}$			& 0.03		& 10.04$^{10.08}_{10.00}$	& 720$^{620}_{830} $		& 0.2\\
Region2		& 10.88$^{10.95}_{10.82}$	& 15$^{10}_{20}$		& 0.004		& 10.17$^{10.19}_{10.14}$	& 970$^{930}_{1100}$		& 0.2		&  9.87$^{9.90}_{9.83}$		& 1900$^{1700}_{2200}$	& 0.5\\
Region3		& 10.60$^{10.65}_{10.55}$	& 60$^{50}_{70}$		& 0.01		& 9.90$^{9.92}_{9.88}$		& 3200$^{3000}_{3300}$		& 0.8		&  9.62$^{9.65}_{9.59}$		& 6900$^{6200}_{7700}$	& 1.7\\
Region4		& 10.17$^{10.22}_{10.12}$	& 230$^{190}_{290}$	& 0.1		& 9.55$^{9.57}_{9.53}$		& 7500$^{7200}_{7900}$		& 1.9		&  9.48$^{9.52}_{9.43}$		& 7100$^{6100}_{8700}$ & 1.8\\
Region5 	& 10.91$^{11.00}_{10.82}$	& 7$^{5}_{11}$		& 0.002		& 10.39$^{10.45}_{10.34}$	& 250$^{200}_{300}$			& 0.1		&  9.71$^{9.75}_{9.67}$		& 2300$^{2000}_{2600}$	& 0.6 \\ 

\hline
\end{tabular}
\tablefoot{The box labelled as BG in the top panel of Fig.~\ref{em_loci} was considered for background subtraction. The upper and lower case numbers indicate the errors.}
\end{table*}
%%------------------------------------------------------------------------

Table.~\ref{table:fill} shows electron densities, filling factors 
and the column depths for five different regions shown in the top panel 
in Fig.~\ref{em_loci} using the ions \ion{Fe}{xii}, \ion{Fe}{xiii} and 
\ion{Fe}{xiv} before and after background subtraction. We have used the 
region labelled 'BG' in the top panel of Fig.~\ref{em_loci} for 
background/foreground subtraction. The filling factor is estimated using 
equation~\ref{fill}. The column depth is estimated using 
equation~\ref{coldepth} assuming a filling factor equal to 1. The
  table clearly demonstrates the importance of background/foreground
  in the measurements of electron densities and filling factors. The
  electron densities for each ion have increased substantially after
  subtracting the background, and the filling factors and column depths have
  substantially decreased. This is the first time the importance of
  background/foreground emission has been demonstrated while estimating physical 
parameters such as density, filling factors and column
  depth in moss regions. After the background subtraction, we find
  substantial increases in the electron densities and meaningful results
  for filling factors. The filling factors derived for \ion{Fe}{xii}
  are very low i.e., much less than 1, whereas those for
  \ion{Fe}{xiii} and \ion{Fe}{xiv} are closer to 1, sometimes even
  more than 1. A filling factor greater than 1 does not give any
  meaningful information. However, in the present case it suggests
  that we have very likely underestimated the column depth by using
  the thickness of the moss measured using TRACE observations.

  The column depth measurements presented in Table~\ref{table:fill},
  which are based on the assumption that the filling factor is 1, show that the moss
  seen in \ion{Fe}{xii} is a very thin region i.e., about the order of
  a fews tens of kilometers in the dense moss regions. At higher
  temperatures e.g., in \ion{Fe}{xiii} and \ion{Fe}{xiv}, the
  estimated column depth is larger than that estimated by
  \ion{Fe}{xii}, by a large factor of $\approx$10-20.  One possible reason for this difference could be that the
  background/foreground is not completely removed. This could explain
  the higher filling factor and larger path length obtained for
  \ion{Fe}{xiii} and \ion{Fe}{xiv}. However, the question remains as
  to why we have such a small column depth for \ion{Fe}{xii} in
  comparison to that which is measured from the limb observations. We note that 
  the densities observed using \ion{Fe}{xii}
  are too high in comparison to those derived using \ion{Fe}{xiii} and \ion{Fe}{xiv} and 
  this is the most likely the reason for very low column
  depth. However, this complex issue 
  involving atomic physics calculations needs further
  investigation in 
  order to understand the discrepancies in densities, filling factors and
  column depth between \ion{Fe}{xii} and other ions.

It is worthwhile emphasizing here that column depths and filling factors are derived 
using the coronal abundances of \cite{coronal_abund}. These values are significantly 
different (a factor of $\sim$3-4 larger) when photospheric abundences are used.

%This difference in the estimated column depth decreases when account is taken of  
%the background/foreground emission. However, the difference still persists.

%%-------------
\section{Summary and Conclusions} \label{con}

Using Hinode/EIS observations, we have studied basic physical plasma parameters such as temperature, electron density, filling factors, and column depth in moss regions and the variation of these parameters over an hour and over a time period of five days. In addition, we have revisited the question of whether the moss regions are the footpoints of hot loops using observations from TRACE, EIS and XRT. The results are summarized below.

\begin{itemize}

\item Based on the TRACE, EIS and XRT observations we find that most of the moss regions are essentially located at the footpoints of hot loops. In some places we observed TRACE~171 (1~MK) loops rooted in the moss regions.

\item Based on the line intensity ratios of \ion{Fe}{xi}~$\lambda$188 and \ion{Fe}{xiii}~$\lambda$202, and an emission measure analysis, we find that the characteristic temperature of moss regions is about log~T~=~6.2. Emission measure analyses over a time period of one hour (Fig.~\ref{emission_measure}) and over five days (Fig.~\ref{five_days_em}) reveal that the thermal structure of the moss regions does not change significantly with time.
  
\item The electron densities measured using \ion{Fe}{xii} ratios are about 1-3~$\times$~10$^{10}$~cm$^{-3}$ and about 2-4~$\times$~10$^{9}$~cm$^{-3}$ using \ion{Fe}{xiii} 
and \ion{Fe}{xiv}. Work is in progress to try to resolve this discrepancy. It is worth emphasizing here that if the \ion{Fe}{xii}~$\lambda$186.8 were blended and we lower its intensity by 20\%, then the electron densities obtained using \ion{Fe}{xii}  would become consistent with those obtained using \ion{Fe}{xiii} and \ion{Fe}{xiv}. The densities derived using \ion{Fe}{xiii} and \ion{Fe}{xiv} are similar to those derived by \cite{fletcher_moss} using \ion{Si}{x} line ratios observed by CDS. However, when we subtract the foreground/background emission we find a substantial increase (a factor of 3-4 or even more in some cases) in the densities.

\item The electron densities only show small changes ($\sim$25\%) over a period of an hour. There are large variations (an order of magnitude increase) in \ion{Fe}{xii} densities when measured over a period of five days. However, the variation in the densities obtained using \ion{Fe}{xiii} and \ion{Fe}{xiv} is only about 50-70\%. 

\item The filling factor of the moss plasma is in the range 0.1-1 and the path length along which the emission originates is from a few 100 to a few 1000 kms long.
   
\end{itemize}

These new measurements of the thermal and density structure in moss regions should provide important constraints for the modelling of loops in the hot and dense core of active regions.

%%-----------------------------------------------------------------------
\section{Acknowledgements} We thank the referee for the
  constructive and thoughtful comments. DT, HEM and GDZ acknowledge the support
from STFC. We thank Brendan O'Dwyer for various discussions. 
Hinode is a Japanese mission developed and
launched by ISAS/JAXA, collaborating with NAOJ as a domestic partner,
NASA and STFC (UK) as international partners. Scientific operation of
the Hinode mission is conducted by the Hinode science team organized
at ISAS/JAXA. This team mainly consists of scientists from institutes
in the partner countries. Support for the post-launch operation is
provided by JAXA and NAOJ (Japan), STFC (U.K.), NASA, ESA, and NSC
(Norway). The help and support of the Hinode/EIS team in particular is
acknowledged.
%%------------------------------------------------------------------------

\bibliographystyle{aa}
\bibliography{new_ref}

\end{document}